\documentclass[review]{elsarticle}

\biboptions{authoryear}

\usepackage{natbib}

\usepackage{setspace}
\usepackage{mathrsfs}
\usepackage[top=1in, bottom=1in, left=1in, right=1in]{geometry}
\usepackage{amsmath}
\usepackage{amsthm}
\usepackage{amssymb}
\usepackage{graphicx}
\usepackage{dsfont} 

\newcommand{\od}[2]{\frac{\text{d} #1}{\text{d} #2}}

\newcommand{\dx}[1]{\ \text{d} #1}
\newcommand{\E}{\mathbb{E}}
\newcommand{\Var}{\text{Var}}

\newcommand{\indicator}[1]{\mathds{1}\left\{ #1 \right\}}

\newtheorem{thm}{Theorem}
\newtheorem{cor}{Corollary}
\newtheorem{lem}{Lemma}

\begin{document}

\begin{frontmatter}
  
\title{On the distribution of interspecies correlation \\
for Markov models of character evolution on Yule trees}

\author[whm]{Willem H. Mulder}
\ead{willem.mulder@uwimona.edu.jm}
\address[whm]{Department of Chemistry, The University of the West Indies, Mona Campus, Kingston 7, Jamaica, West Indies}

\author[fwc]{Forrest W. Crawford}
\ead{forrest.crawford@yale.edu}
\address[fwc]{Department of Biostatistics, Yale School of Public Health, New Haven, CT, USA}

\begin{abstract}
Efforts to reconstruct phylogenetic trees and understand evolutionary processes depend fundamentally on stochastic models of speciation and mutation.  The simplest continuous-time model for speciation in phylogenetic trees is the Yule process, in which new species are ``born'' from existing lineages at a constant rate.  Recent work has illuminated some of the structural properties of Yule trees, but it remains mostly unknown how these properties affect sequence and trait patterns observed at the tips of the phylogenetic tree. Understanding the interplay between speciation and mutation under simple models of evolution is essential for deriving valid phylogenetic inference methods and gives insight into the optimal design of phylogenetic studies.  In this work, we derive the probability distribution of interspecies covariance under Brownian motion and Ornstein-Uhlenbeck models of phenotypic change on a Yule tree.  We compute the probability distribution of the number of mutations shared between two randomly chosen taxa in a Yule tree under discrete Markov mutation models.  Our results suggest summary measures of phylogenetic information content, illuminate the correlation between site patterns in sequences or traits of related organisms, and provide heuristics for experimental design and reconstruction of phylogenetic trees. 
\end{abstract}

\end{frontmatter}


\section{Introduction}

Simple stochastic models of speciation and trait evolution have proven useful for reconstruction of phylogenetic trees describing the ancestral relationship between sets of taxa.   The simplest continuous-time model of speciation is the Yule process, in which each extant lineage gives birth at constant rate $\lambda$.  A Yule tree is a phylogenetic tree in which the branching times of the tree are drawn from the Yule distribution.   Despite the apparent simplicity of the Yule process, Yule trees have complex structural properties \citep{Steel2002Shape,Rosenberg2006Mean,Gernhard2008Stochastic,Steel2010Expected,Mulder2011Probability,Crawford2013Diversity}.  The Yule process is usually employed as a prior or null distribution on the space of phylogenetic trees within a broader scheme of phylogenetic reconstruction \citep{Nee1994Reconstructed,Rannala1996Probability,Nee2006Birth}.  Researchers impose a model for the evolution of a character (trait, DNA, RNA, or amino acid sequence) on the branches of this phylogenetic tree.  By jointly estimating the phylogenetic tree topology, branch lengths, and the parameters underlying the evolutionary model, researchers hope to understand the evolutionary history and process that gave rise to the observed data.  



Research on the interaction of tree topology, branch lengths, and evolutionary processes generally falls into one of two categories.  The first is the search for better measures of phylogenetic information for \emph{prospective} experimental design.  Most of these studies examine the probability of correctly reconstructing a simple tree or optimal design of phylogenetic studies \citep{Yang1998Best,Sullivan1999Effect,Shpak2000Information,Zwickl2002Increased,Susko2002Testing}.  Several authors have attempted to determine whether it is better to add more taxa or additional characters to maximize the chance of reconstructing the correct tree \citep{Graybeal1998Better,Zwickl2002Increased}.  \citet{Steel2000Parsimony} analyze basic models of evolution to understand the theoretical properties of stochastic models on phylogenetic trees.
\citet{Fischer2009Sequence} consider asymptotic sequence length bounds for correct reconstruction under maximum parsimony.  \citet{Townsend2007Profiling} introduces ``phylogenetic informativeness'', the probability of observing site patterns allowing correct reconstruction of a four-taxon tree.  \citet{Susko2011Large} and \citet{Susko2012Probability} find expressions for correct reconstruction probability for small internal edges on four-taxon trees.  Real-world phylogenetic studies often involve large numbers of taxa, and it remains controversial whether properties of mutation models on four-taxon trees generalize to trees with larger numbers of taxa \citep[see e.g.][]{Townsend2007Profiling,Klopfstein2010Evaluation,Townsend2011Taxon}.

The second class of approaches focuses on \emph{retrospective} inferences about evolutionary parameters and the derivation of estimators and confidence intervals.  Following the work of \citet{Stadler2009Incomplete}, who describes sampling properties of birth-death trees and the distribution of the age of the most recent common ancestor (MRCA) of subsets of randomly chosen taxa, \citet{Bartoszek2012Phylogenetic} and \citet{Bartoszek2013Quantifying} find expressions for the expectation of the interspecies correlation under models of continuous trait evolution via diffusion and Ornstein-Uhlenbeck processes. \citet{Bartoszek2012Phylogenetic} derive asymptotic confidence intervals for ancestral trait values under these models.  \citet{Crawford2013Diversity} give an estimator for the evolutionary variance under Brownian motion for an unobserved Yule tree.

In this paper we study the distribution of character values observed at the tips of a phylogenetic tree generated by the Yule process.  We first state two theorems that describe the distribution of the time of shared ancestry between two randomly chosen taxa in a Yule tree of age $\tau$ with $n$ taxa and speciation rate $\lambda$.  
Next we extend results presented by \citet{Bartoszek2012Phylogenetic} and \citet{Bartoszek2013Quantifying} to find the exact probability distribution and covariance between pairs of randomly chosen tip values under Brownian motion and Ornstein-Uhlenbeck evolution of a continuous trait.  These results give insight into the finite-time, finite-$n$ dynamics of interspecies correlation.  Next we examine discrete character evolution on Yule trees under Poisson and reversible Poisson mutation models.  
We suggest a new measure of phylogenetic information and give a method for deciding whether it is better to add taxa or sites to a phylogenetic analysis.  

\section{Background}

A Yule process $Y(t)$ is a continuous-time Markov chain on the positive integers in which a jump from state $n$ to $n+1$ occurs with rate $n\lambda$.  Define $P_{mn}^Y(t)=\Pr(Y(t)=n\mid Y(0)=m)$ to be the transition probability from state $m$ to $n$ in time $t$.  The Yule process obeys the forward Kolmogorov equations
\begin{equation}
  \od{P_{mn}^Y(t)}{t} = (n-1)\lambda P^Y_{m,n-1}(t) - n\lambda P^Y_{mn}(t) 
  \label{eq:fwd}
\end{equation}
for $m\geq 1$ and $n\geq m$. The transition probabilities are 
\begin{equation}
  P_{mn}^Y(t) = \binom{n-1}{m-1} e^{-\lambda m t} (1-e^{-\lambda t})^{n-m}
  \label{eq:yuleprob}
\end{equation}
\citep{Bailey1964Elements}.  A Yule tree is a binary tree in which the number of extant lineages at time $t$ is given by the Yule process $Y(t)$.  If there are $n$ extant lineages and a ``birth'' event occurs, one of the $n$ lineages is chosen uniformly at random and split into two.  
In this paper, we assume that at the MRCA of all $n$ taxa existed at time 0. We model $t=0$ as the time of the first split, so $Y(0)=2$, and both tree size (number of taxa) $n$ and age $\tau$ are given. In what follows, we limit our attention to the $(n-1)!$ unlabelled, ranked, oriented trees that make up an $n$-forest, since our conclusions readily carry over to the $n!(n-1)!/2^{n-1}$ leaf-labelled, ranked Yule trees of phylogenetic interest \citep{Gernhard2008Stochastic,Mulder2011Probability}.  




\begin{figure}
  \centering
  \includegraphics[width=0.6\textwidth]{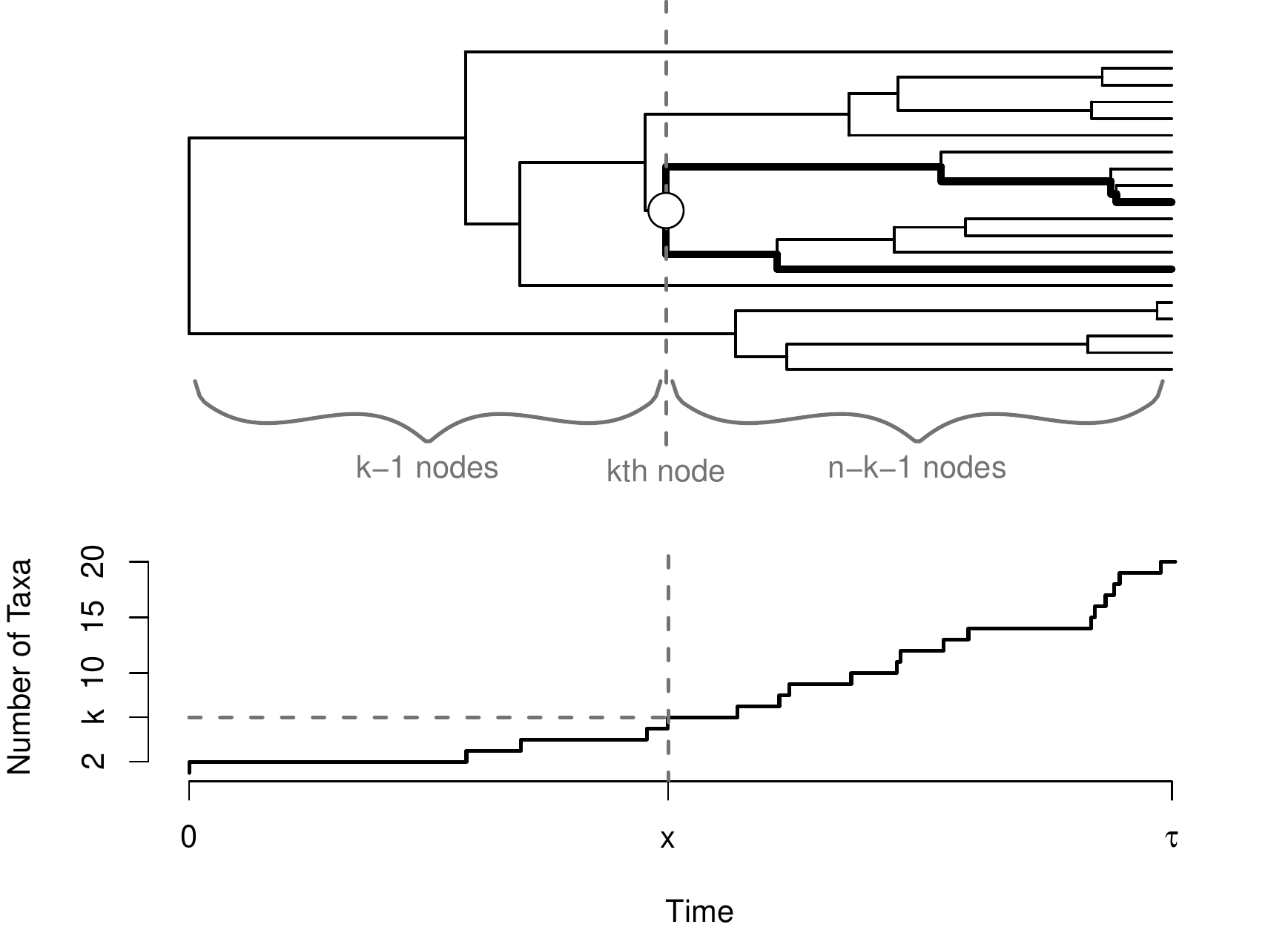}
  \caption{The most recent common ancestor (MRCA) of two taxa at time $x$ in a Yule tree and the corresponding Yule counting process.  The Yule tree in the top panel has $Y(0)=2$ and $Y(\tau)=n$.  The lineages connecting two randomly chosen taxa to their MRCA at the $k$th node are shown as thick lines.  The MRCA is shown as a circle.  The bottom panel shows the corresponding Yule process counting the number of lineages over time.  The time of the birth of the $k$th lineage $x$ is shown with a dashed vertical line.}
  \label{fig:treefig}
\end{figure}

We now consider pairs of tips on a Yule tree whose MRCA is the $k$th birth event.  We call these events ``nodes'' in the tree. The $k$th node is preceded chronologically by $k-1$ nodes, and this node emerges at time $x$ since the first split.  The $k$th node corresponds to the ``crown age'' of the sub-tree or clade below the node.  Figure \ref{fig:treefig} shows an example in which the $k$th birth event, preceded by $k-1$ such events, takes place at time $x$.  In continuous time each $n$-tree pattern of this type, with tree age $\tau$ and with the $k$th node appearing at time $x$, has the same probabilistic weight,
and hence these trees can be dealt with on equal footing using purely combinatorial arguments.  The following result gives the probability of two randomly chosen tips in a phylogenetic tree having their MRCA at the $k$th node.  It was first derived by \citet{Stadler2009Incomplete}.
\begin{thm}
\label{thm:Pnk}
The probability of randomly choosing two tips in a tree of size $n$ whose MRCA is the $k$th node is
\begin{equation}
  P(n,k) = \frac{2(n+1)}{(n-1)(k+1)(k+2)} 
\end{equation}
for $n\geq k+1$ \citep{Stadler2009Incomplete}.
\end{thm}
\noindent \ref{app:Pnk} gives a simple alternative proof of this fact using recurrence relations.



We now consider the time of shared ancestry of two randomly chosen taxa, the age of their MRCA.  Theorem \ref{thm:Pnk} provides the probability of choosing two tips whose MRCA is the $k$th node; here we seek the distribution of the age $x$ of this node. 
\begin{lem}
\label{lem:px}
The probability density of the time $x$ of the $k$th node of a Yule tree of age $\tau$ and size $n$ is 
\begin{equation}
  p(x|k,n,\tau,\lambda) = \begin{cases}
    \delta(x) & k=1 \\[1em]
              \frac{\displaystyle \lambda (n-2)\binom{n-3}{k-2} e^{-(k-1)\lambda(\tau-x)} (1-e^{-\lambda x})^{k-2} (1-e^{-\lambda(\tau-x)})^{n-k-1}}{\displaystyle (1-e^{-\lambda \tau})^{n-2}} & k\geq 2 
  \end{cases}
  \label{eq:px}
\end{equation}
for $0\leq x \leq \tau$, where $\delta(x)$ is the Dirac delta function.
\end{lem}
\noindent \ref{app:px} provides a derivation.

Now we study the age of the MRCA of two randomly chosen taxa without conditioning on the MRCA being the $k$th node in the tree.  Finding the marginal distribution of $x$ by summing $P(n,k)$ over $k$ with respect to \eqref{eq:px}, we arrive at 
\begin{equation}
 p(x|n,\tau,\lambda) = \sum_{k=1}^{n-1} P(n,k)\ p(x|k,n,\tau,\lambda) .
 \label{eq:pxn}
\end{equation}
where $P(n,k)$ is given by Theorem \ref{thm:Pnk} and $p(x|k,n,\tau,\lambda)$ is given by Lemma \ref{lem:px}.  The following Theorem gives a closed-form expression for this probability.
\begin{thm}
\label{thm:succinct}
The probability density of the age $x$ of the MRCA of two randomly chosen taxa in a tree of size $n$ and age $\tau$ with branching rate $\lambda$ is
\begin{equation}
  \begin{split}
    p(x|n,\tau,\lambda) &= \frac{n+1}{3(n-1)}\delta(x) 
    + \frac{2\lambda (e^{\lambda(\tau-x)}-1)^3 (1-e^{-\lambda\tau})}{n(n-1)^2 (1-e^{-\lambda x})^4 (1-e^{-\lambda(\tau-x)})} 
    \Bigg[ (n(n-3)+2)\left(\frac{1-e^{-\lambda x}}{e^{\lambda(\tau-x)}-1}\right)^2  \\
      &\quad - 4(n-2)\frac{1-e^{-\lambda x}}{e^{\lambda(\tau-x)}-1} 
    - 2\left((n+1)\frac{1-e^{-\lambda x}}{e^{\lambda(\tau-x)}-1} + 3 \right) 
  \left(\frac{1-e^{-\lambda (\tau-x)}}{1 - e^{-\lambda\tau}} \right)^{n-1} + 6 \Bigg] ,
  \end{split}
\label{eq:fullsuccinct}
\end{equation}
where $\delta(x)$ is the Dirac delta function.
\end{thm}
\noindent \ref{app:succinct} gives a proof.  To our knowledge, this is a new result; \citet{Stadler2009Incomplete} gives a similar derivation for the distribution of the MRCA age of two randomly picked taxa in a Yule tree, but without conditioning on $\tau$.


\section{Markov models of evolutionary change}

\begin{figure}
  \centering
  \includegraphics[width=\textwidth]{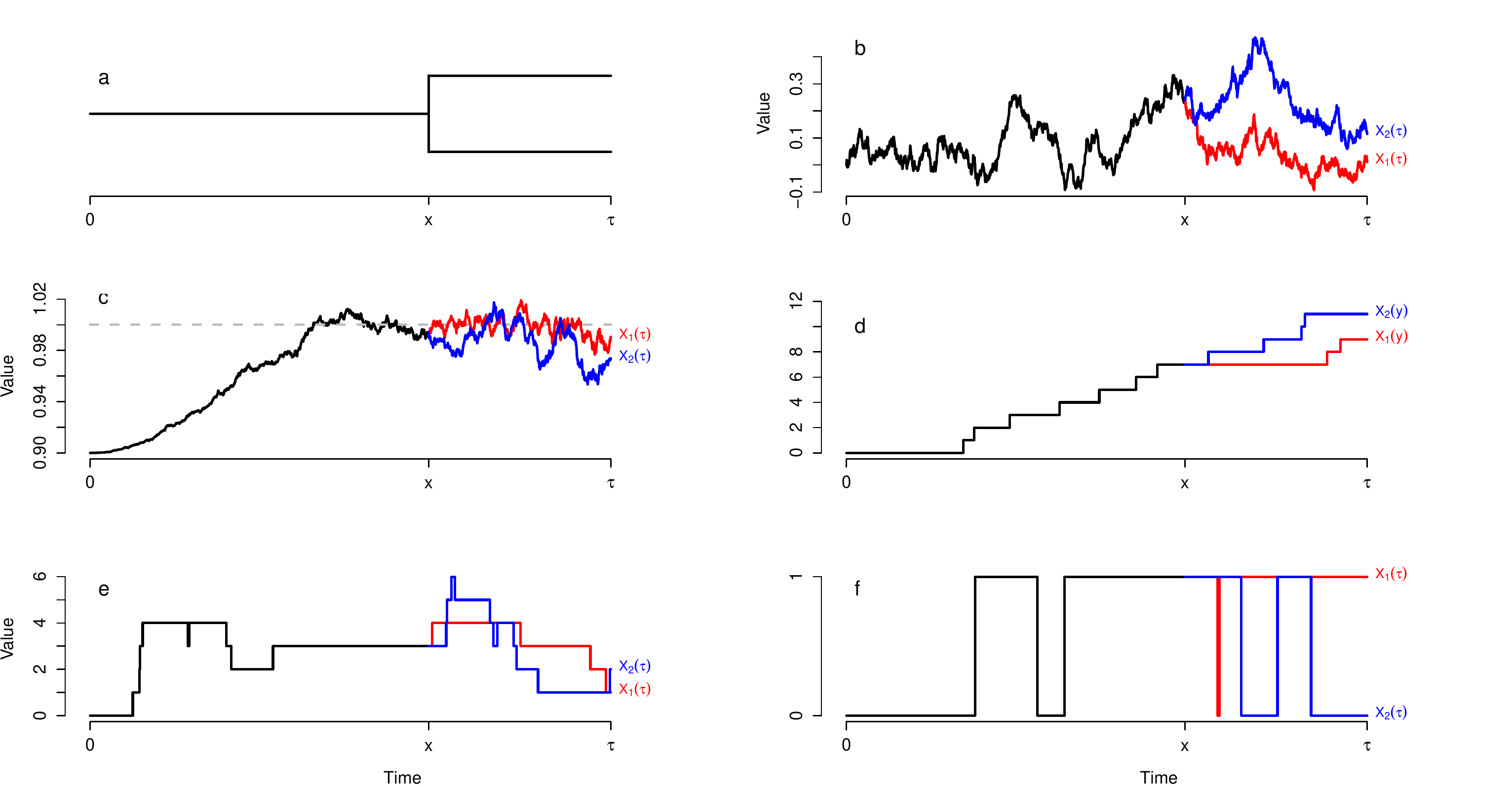}
  \caption{Examples of Markov models on a two-taxon tree of age $\tau=2$ and splitting time $x=1.3$.  Panel (a) shows a Yule tree for two taxa with a splitting time at $x$; (b) shows a Brownian motion model with $\mu=$ and $\sigma^2=10^{-4}$; (c) Ornstein-Uhlenbeck with $\theta=1$, $\sigma^2=10^{-4}$, and $\alpha=0.005$; (d) Poisson with $\alpha=4$, (e) reversible Poisson with $\alpha=4$ and $\beta=2$, and (f) binary with $\alpha=2$ and $\beta=2$.  The correlation between the character values $X_1(\tau)$ and $X_2(\tau)$ is a function of their time of shared ancestry $x$.  }
  \label{fig:markovexamples}
\end{figure}

Now consider a Markov process $X(t)$ on the branches of a Yule tree of age $\tau$.  
Figure \ref{fig:markovexamples} shows examples of Markov models $X(t)$ that we will consider.  In the panel (a), a Yule tree of age $\tau$ and size $n=2$ taxa is shown with a speciation event at time $x$.  The process begins on the ancestral lineage and evolves over the interval $(0,x)$.  Following the speciation event at time $x$, the character evolves \emph{independently} in the two daughter taxa, conditional on their shared ancestral trait value $X(x)$.  Next are the examples of the Markov evolutionary models $X(t)$ that we study in this paper: Brownian motion (b), Ornstein-Uhlenbeck process (c), Poisson (d), Reversible Poisson (e), and Binary (f).  Parameters used to generate these simulated trajectories are given in the caption of Figure \ref{fig:markovexamples}.

In each of the following sections, we will consider a statistic $M$, which is a function of the value of $X_i(\tau)$ and $X_j(\tau)$ at two randomly chosen tips $i$ and $j$, with $i\neq j$.  For example, $M$ could be the number of shared mutations at the tips.  Then the probability density of $M$ (or mass function if $M$ is discrete-valued) $p(m|n,\tau,\lambda)$ is obtained by marginalizing over the time of shared ancestry,
\begin{equation}
p(m|n,\tau,\lambda) = \sum_{k=1}^{n-1} P(n,k) \int_0^\tau p(x|k,n,\tau,\lambda)\ p(m|x,\tau) \dx{x} ,
\label{eq:PrM}
\end{equation}
where $p(m|x,\tau)$ is the density (or mass function) of $M$, conditional on the time $x$ of shared ancestry.  The following sections describe specific continuous-time Markov models $X(t)$ and statistics $M$ that are of interest in phylogenetic reconstruction and evolutionary inference.


\subsection{Brownian motion}

\label{sec:bm}

Consider a continuous-valued Markov process $X(t)$ with mean zero and variance $\sigma^2$ obeying the stochastic differential equation
$\dx{X(t)} = \sigma \dx{B(t)}$,
where $B(t)$ is Brownian noise.  The process evolves on the branches of a Yule tree.  On a single branch, $X(t)$ has normally distributed increments
$
  X(t) - X(s) \sim \text{Normal}\big(0,\sigma^2(t-s)\big) 
$
for $0\leq s\leq t$.  Given a tree topology $\mathcal{T}$ and branch lengths $\mathbf{t} = (t_1,\ldots,t_{n-1})$, the trait values at the tips of the tree are distributed according to the multivariate random variable
$
  \mathbf{X} \sim \text{Normal}\big(\mathbf{0}, \sigma^2 \mathbf{C}(\mathcal{T},\mathbf{t})\big)
  $
where the evolutionary covariance matrix $\mathbf{C}(\mathcal{T},\mathbf{t})$ is an $n\times n$ matrix whose diagonal elements are equal to $\tau$.  The off-diagonal element $\mathbf{C}_{ij}$ where $i\neq j$ is the time of shared ancestry of taxa $i$ and $j$, and so the matrix is symmetric.   The covariance of the tip values in taxa $i$ and $j$ is proportional to their time of shared ancestry.  This fact gives a natural way of seeing \eqref{eq:fullsuccinct} as the marginal distribution of a randomly-chosen off-diagonal element of $\mathbf{C}$, as the following Corollary makes clear.
\begin{cor}
  For a Yule tree of size $n$ and age $\tau$, a randomly chosen off-diagonal element of the covariance matrix $\mathbf{C}$ has probability density given by \eqref{eq:fullsuccinct} in Theorem \ref{thm:succinct}.
\end{cor}

\citet{Sagitov2012Interspecies} study the ``interspecies correlation coefficient'' 
\begin{equation}
  \rho_n = \frac{1}{\binom{n}{2}\Var\big(X(\tau)\big)} \sum_{j<i} \text{Cov}\big(X_i(\tau),X_j(\tau)\big)
  \label{eq:sagitov}
\end{equation}
where $\Var\big(X(\tau)\big)$ is the variance of the Brownian process on a single lineage of length $\tau$.  We can understand \eqref{eq:sagitov} as the mean pairwise correlation between tip values.  An alternative perspective arises if we instead use \eqref{eq:PrM} to study the distribution of the ``marginal'' pairwise correlation coefficient.  We condition on the tree age $\tau$, so $\Var\big(X(\tau)\big)=\sigma^2\tau$.  Let $M_{ij} = \text{Cov}\big(X_i(\tau),X_j(\tau)\big)$ where $i$ and $j$ are chosen randomly and define $\tilde{\rho}_n = M_{ij}/(\tau\sigma^2)$.  By the change of variables formula, $\tilde{\rho}_n$ has density
\begin{equation}
  p(\tilde{\rho}_n|n,\tau,\lambda) = \tau\sigma^2 p(\tilde{\rho}_n\tau\sigma^2| n,\tau,\lambda)
 \label{eq:bmcorr}
\end{equation}
where the density on the right-hand side of \eqref{eq:bmcorr} is given by \eqref{eq:fullsuccinct}.  This result gives a different perspective on the expressions derived by \citet{Sagitov2012Interspecies} because it provides a distribution for $\tilde{\rho}_n$, rather than an expectation, giving insight into the distribution of correlation between taxa across random pairs of taxa. 



\begin{figure}
  \centering
  \includegraphics[width=0.5\textwidth]{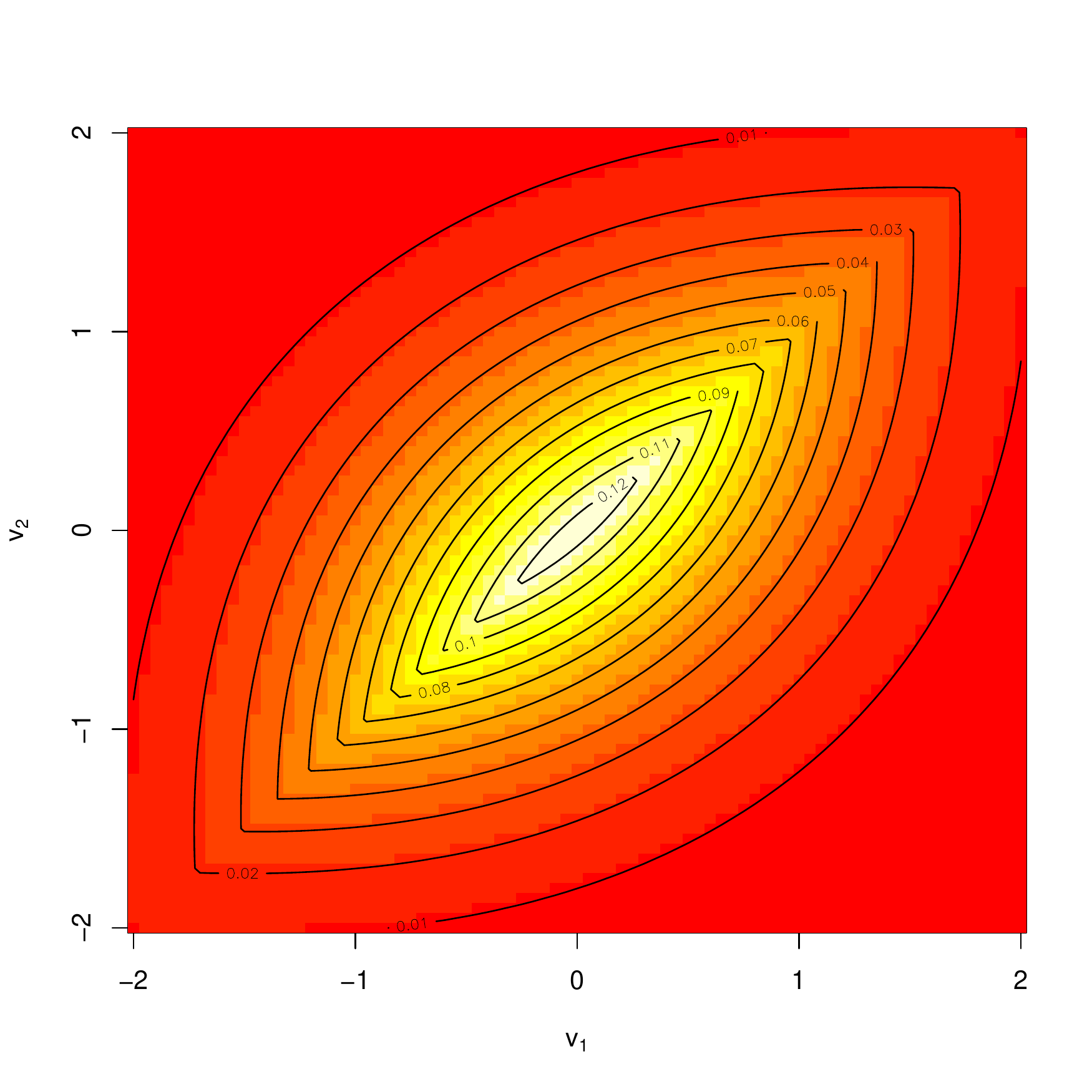}
  \caption{ Joint distribution of trait values at two randomly chosen tips in a tree of size $n=6$ and age $\tau=1$ under the Brownian motion model with zero mean and variance $\sigma^2=1$.  Note that this distribution is \emph{not} bivariate normal.}
  \label{fig:browniancov}
\end{figure}

Furthermore, the ``marginal'' joint distribution of $\big(X_i(\tau),X_j(\tau)\big)$ can be recovered by letting  
\begin{equation}
  \mathbf{C} = \begin{pmatrix} 
          \tau & x    \\
          x    & \tau \end{pmatrix} 
  \end{equation}
 be the evolutionary correlation matrix induced by a two-taxon tree with splitting time $x$.  Let $v=(v_1,v_2)$ be the trait values at the chosen tips.  The marginal distribution of $v$ for two randomly chosen taxa in a tree of size $n$ and age $\tau$ is given by
    \begin{equation}
  f(v_1,v_2) = 
             \sum_{k=1}^{n-1} P(n,k) \int_0^\tau p(x|k,n,\tau,\lambda) \frac{1}{2\pi\sigma^2\sqrt{\tau^2-x^2}}\exp\left[ -\frac{1}{2\sigma^2}\frac{\tau(v_1^2+v_2^2) - 2v_1v_2 x}{\tau^2-x^2}\right]\dx{x} \\
\end{equation}
where ancestral state at time 0 is $(0,0)'$.  While there is little hope of solving the above expression analytically, it is straightforward to evaluate by numerical integration.  Figure \ref{fig:browniancov} shows the joint distribution of $(v_1,v_2)$ for $n=10$ and $\sigma^2=1$.  This joint probability distribution, induced by random choice of two tips, is not bivariate normal.
    
\subsection{Ornstein-Uhlenbeck process}

The Ornstein-Uhlenbeck (OU) process is an extension of the Brownian motion model presented above.  In evolutionary inference, it is used to model selection of a continuous-valued trait toward a global optimum \citep{Butler2004Phylogenetic}.  The OU process obeys the stochastic differential equation
\begin{equation}
  \text{d}X(t) = \alpha[\theta - X(t)]\text{d}t  + \sigma \text{d}B(t) 
\end{equation}
where $\alpha$ is the strength of selection, $\theta$ is the selective optimum, and $\sigma^2$ is the variance of the Brownian noise.  The model gives 
\begin{equation}
  \begin{split}
    \E[X(t)\mid X(0)=\theta_0] &= \theta - (\theta-\theta_0) e^{-\alpha t} \\
  \text{Var}[X(t)\mid X(0)=\theta_0] &= \frac{\sigma^2}{2\alpha} (1-e^{-2\alpha t}) .
  \end{split}
\end{equation}
Now consider two taxa $i$ and $j$ whose MRCA is at time $x$ in a tree of age $\tau$.  As before, let $M_{ij} = \text{Cov}\big(X_i(\tau), X_j(\tau)\big)$ where $i$ and $j$ are chosen randomly from $n$ tips in a tree of age $\tau$, and $i\neq j$.  

Under the OU process, the covariance of the values at $i$ and $j$ is given by
\begin{equation}
  M_{ij}(x) = \frac{\sigma^2}{2\alpha} e^{-2\alpha(\tau-x)}(1-e^{-2\alpha x}) .
\end{equation}
where $x$ is the age of the MRCA of $i$ and $j$.  Solving for $x$ gives 
\begin{equation}
  x(m) = \frac{1}{2\alpha} \log\left[ 1 + \frac{2\alpha e^{2\alpha\tau}}{\sigma^2} m \right].
  \label{eq:oux}
\end{equation}
where $x(m)$ is interpreted as a function of $M_{ij}=m$.  Applying the change of variables formula, we find that the distribution of the covariance between two randomly chosen tips under the OU process is
\begin{equation}
  p(m|n,\tau,\lambda,\alpha,\sigma^2) = p(x(m)|n,\tau,\lambda) \frac{e^{2\alpha\tau}}{\sigma^2 + 2\alpha e^{2\alpha \tau}m} .
  \label{eq:oucov}
\end{equation}
where the density on the right-hand side of \eqref{eq:oucov} is given by \eqref{eq:fullsuccinct}.  

Now we find the distribution of the interspecies correlation coefficient in a similar way as above by letting
$
  \tilde{\rho}_n = M_{ij}/\text{Var}\big(X_i(\tau)\big) .
  $
Then the density of $\tilde{\rho}_n$ is 
\begin{equation}
  p(\tilde{\rho}_n|n,\tau,\lambda) = p(m(\tilde{\rho}_n)|n,\tau,\lambda,\alpha,\sigma^2) \frac{\sigma^2}{2\alpha}(1-e^{-2\alpha\tau}) .
\end{equation}


\subsection{Poisson}

\label{sec:poi}

Suppose there are infinitely many sites in a DNA sequence, and a mutation occurs at some site with rate $\alpha$ \citep{Durrett2008Probability}.  We assume that the mutations are irreversible, unique, and distinguishable when observed at the tips of the tree, so we can view the accumulation of mutations on a phylogenetic tree as a Poisson process $X(t)$ counting the number of mutations up to time $\tau$, the age of the tree.  Let $P_k(t) = \Pr(X(t)=k|X(0)=0)$.  The forward equations for this process are
\begin{equation}
\od{P_k(t)}{t} = \alpha P_{k-1}(t) - \alpha P_k(t) .
\end{equation}
Solving for $P_k(t)$ with the initial condition $P_0(0)=1$, we arrive at the familiar Poisson transition probability $P_k(t) = (\alpha t)^k e^{-\alpha t}/k!$ .

Before stating the main result for this section, we provide an intermediate Corollary describing the relationship between Poisson mutations and the total branch length (also known as phylogenetic diversity) of a Yule tree: if Poisson mutations fall on a Yule tree on $n$ tips, age $\tau$, branching rate $\lambda$ with rate $\alpha$, then straightforward application of a result from \citet{Crawford2013Diversity} gives the following result.
\begin{cor}
Suppose Poisson mutations with rate $\alpha$ fall on a Yule tree of size $n$, age $\tau$, and branching rate $\lambda$.  Let $M$ be the number of mutation events on the tree.  Then 
\begin{equation}
M \sim \text{Poisson}\left(\alpha \int_0^\tau Y(x)\dx{x}\right) 
\end{equation}
and the probability mass function of $M$ is given by
\begin{equation}
  \begin{split}
  \Pr(M = m) &= \frac{\alpha^m\lambda^{n-2}}{m!(n-3)P^Y_{2n}(\tau)} \sum_{j=2}^n \binom{n-1}{j-1} (j-1) (-1)^{j} \sum_{\ell=0}^{n-3}\frac{\binom{n-3}{\ell}(-j\tau)^{n-\ell-3}}{(\alpha+\lambda)^{m+\ell}} \\
  &\qquad\qquad \times \left[ \gamma\big(m+\ell+1,n\tau(\alpha+\lambda)\big) - \gamma\big(m+\ell+1,j\tau(\alpha+\lambda)\big)\right]
\end{split}
\end{equation}
where $\gamma(\cdot,\cdot)$ is the lower incomplete gamma function and $P^Y_{2n}(\tau)$ is the Yule transition probability.
\label{cor:pMpoisson}
\end{cor}
\noindent A proof is given in \ref{app:pMpoisson}.  The quantity $M$ is also known as the ``number of segregating sites'' on the tree.


Now let $M$ be the number of Poisson mutations shared by two randomly chosen taxa in a Yule tree of age $\tau$ and size $n$.  The probability of the two taxa sharing $m$ mutations is the probability that $m$ mutations accumulated during the time of their shared ancestry. If the time of shared ancestry is $x$, then the number of mutations occurring during this time is Poisson with rate $\alpha x$. Then marginalizing over $x$, we recover the distribution of the number of shared mutations between two randomly chosen taxa.  When $M=0$, 
\begin{equation}
  \begin{split}
    \Pr(M=0|n,\tau,\lambda,\alpha) &= \frac{n+1}{3(n-1)} + \frac{2(n+1)(n-2)}{(n-1)(1-e^{-\lambda\tau})^{n-2}}\sum_{k=2}^{n-1} \binom{n-3}{k-2} \frac{e^{-(k-1)\lambda\tau} }{(k+1)(k+2)} \\
                           &\quad \times \sum_{j=0}^{k-2} \binom{k-2}{j} \sum_{i=0}^{n-k-1} \binom{n-k-1}{i} (-1)^{i+j} e^{-i\lambda\tau} \frac{\gamma\big(m+1,(\alpha+\lambda(j-i-k+1))\tau\big)}{(\alpha+\lambda(j-i-k+1))^{m+1}} 
  \end{split}
  \label{eq:poisPm0}
\end{equation}
and when $M\geq 1$, we have 
\begin{equation}
  \begin{split}
    \Pr(M=m|n,\tau,\lambda,\alpha) &= \frac{2\lambda(n+1)(n-2)\alpha^m}{m!(n-1)(1-e^{-\lambda\tau})^{n-2}}\sum_{k=2}^{n-1} \binom{n-3}{k-2} \frac{e^{-(k-1)\lambda\tau} }{(k+1)(k+2)} \\
                              &\quad \times \sum_{j=0}^{k-2} \binom{k-2}{j} \sum_{i=1} \binom{n-k-1}{i} (-1)^{i+j} e^{-i\lambda\tau} \frac{\gamma\big(m+1,(\alpha+\lambda(j-i-k+1))\tau\big)}{(\alpha+\lambda(j-i-k+1))^{m+1}} 
\end{split}
 \label{eq:poisPm}
\end{equation}
Derivations of \eqref{eq:poisPm0} and \eqref{eq:poisPm} are given in \ref{app:poisPm}.


\subsection{Poisson/reversible}

\label{sec:revpoi}


In this model, unique distinguishable mutations occur as a Poisson process with constant rate $\alpha$, but the changes are reversible with rate $\beta$ per mutation.  When there are $j$ mutations in a lineage, the rate of loss is $j\beta$.  This corresponds to reversion of disadvantageous changes in an evolutionary context.  On a single branch, the number of mutations added and removed is modeled by the $M/M/\infty$ queue, also known as the immigration-death process.  The process has forward equation
\begin{equation}
\od{P_{ij}(t)}{t} = \alpha P_{i,j-1}(t) + (j+1)\beta P_{i,j+1}(t) - (\alpha + j\beta)P_{ij}(t) .
\label{eq:poisrev}
\end{equation}
Let $X(t)$ be the number of mutations at time $x$, given that there were none at time 0.  Then $X(t)$ has Poisson distribution with mean $(\alpha/\beta)(1-e^{-\beta t})$.

Suppose now that $k$ mutations exist at time 0 and we monitor only on the loss of those original $j$ mutations. We are interested in the probability of losing $j-m$ mutations, so that $m$ remain at time $t$.  The forward equations become
\begin{equation}
\od{P_{jm}(t)}{t} = (m+1) \beta P_{j,m+1}(t) - m\beta P_{jm}(t) 
\label{eq:puredeath}
\end{equation}
for $m\leq j$.  On a single branch of length $t$, the number of surviving mutations $m$ is Binomially distributed with probability $e^{-\beta t}$.  The number of mutations $M$ shared between two randomly chosen taxa has probability mass function 
\begin{equation}
    \Pr(M=m|x,\tau,\alpha,\beta) = \frac{\exp\left[-\frac{\alpha}{\beta}(1-e^{-\beta x})e^{-2\beta(\tau-x)} \right] \left[ \frac{\alpha}{\beta}(1-e^{-\beta x}) e^{-2\beta(\tau-x)} \right]^m }{m!}.
  \label{eq:poisrevPm}
\end{equation}
Therefore when the time of shared ancestry is $x$ and tree age is $\tau$, $M$ has Poisson distribution.  A proof of \eqref{eq:poisrevPm} is given in \ref{app:poisrevPm}.  Then the distribution of $M$ is given by 
\begin{equation}
  \Pr(M=m|n,\tau,\lambda,\alpha,\beta) = \sum_{k=1}^{n-1} P(n,k) \int_0^\tau p(x|k,n,\tau,\lambda) \Pr(M=m|x,\tau,\alpha,\beta) \dx{x}
\end{equation}
which does not seem to have a simple closed-form expression.  It can be easily evaluated by numerical integration.


\subsection{Binary characters and identity-by-descent}

\label{sec:binary}

We now study a two-state process $X(t)$ occurring independently at $N$ independent sites in a DNA sequence evolution model.  This model is similar to the Reversible Poisson model, but for a finite number of sites.  Call the two states in the process 0 and 1.  Transitions from 0 to 1 occur with rate $\alpha$ and from 1 to 0 with rate $\beta$.  The meaning of $\alpha$ and $\beta$ here is different than in Section \ref{sec:revpoi}. In this context, $\alpha$ and $\beta$ are per-site mutation rates.  The transition rate matrix is 
\begin{equation}
Q = \begin{pmatrix} -\alpha & \alpha \\ \beta   & -\beta \end{pmatrix}. 
\end{equation}
The system evolves according to the matrix differential equation $\od{P^B}{t} = P^B(t) Q $.  With initial condition $P^B(0) = I$, the transition probability matrix is given by $P^B(t)=e^{Qt}$, where the elements of $P^B(t)$ are 
\begin{align}
  P^B_{00}(t) &= \frac{\beta+\alpha e^{-(\alpha+\beta) t}}{\alpha+\beta} &  P^B_{01}(t)  &= \frac{\alpha}{\alpha+\beta}\left(1-e^{-(\alpha+\beta) t}\right) \nonumber \\
  P^B_{10}(t) &= \frac{\beta}{\alpha+\beta} \left(1 - e^{-(\alpha+\beta)t}\right) &  P^B_{11}(t)  &= \frac{\alpha+\beta e^{-(\alpha+\beta) t}}{\alpha+\beta}.
\label{eq:ibsprobs}
\end{align}

We are interested in the probability that $K$ of $N$ sites share the same state, conditional on the state being inherited  from the MRCA.  That is, we wish to exclude situations in which the taxa show a matching site pattern, but the matching states are not directly descended from the same ancestral state.  Site patterns having this property are called ``identical by descent''.  To illustrate, consider the statistic $M_k=\indicator{X_{i,k}(\tau)=X_{j,k}(\tau),\text{ibd}}$ which is 1 when site $k$ in taxon $i$ and site $k$ in taxon $j$ share the same ancestral allele identically by descent, and zero otherwise.  Then 
\begin{equation}
\Pr(M_k=1|x,\text{ibd}) = P^B_{00}(x) e^{-2\alpha(\tau-x)} + P^B_{01}(x) e^{-2\beta(\tau-x)} 
\end{equation}
where $P^B_{00}(x)$ and $P^B_{01}(x)$ are given in \eqref{eq:ibsprobs}.  

Letting $M=\sum_{k=1}^N M_k$, the probability that $K$ of $N$ sites match IBD is binomial,
\begin{equation}
  \Pr(M = K|x,\text{ibd}) = \binom{N}{K} \Pr(M_k=1|x,\text{ibd})^K \Pr(M_k=0|x,\text{ibd})^{N-K} .
  \label{eq:bink}
\end{equation}
Marginalizing over the time $x$ of shared ancestry,
\begin{equation}
 \Pr(M = K|\text{ibd}) = \sum_{k=1}^{n-1} P(n,k) \int_0^\tau p(x|\tau,n,\lambda) \Pr(M=K|x,\text{ibd}) \dx{x}.
\end{equation}
This model will be useful in the next section, where we consider phylogenetic experimental design.


\section{Applications}

\label{sec:applications}

\subsection{Phylogenetic information and uncertainty}

\begin{figure}
  \centering
  \includegraphics[width=0.5\textwidth]{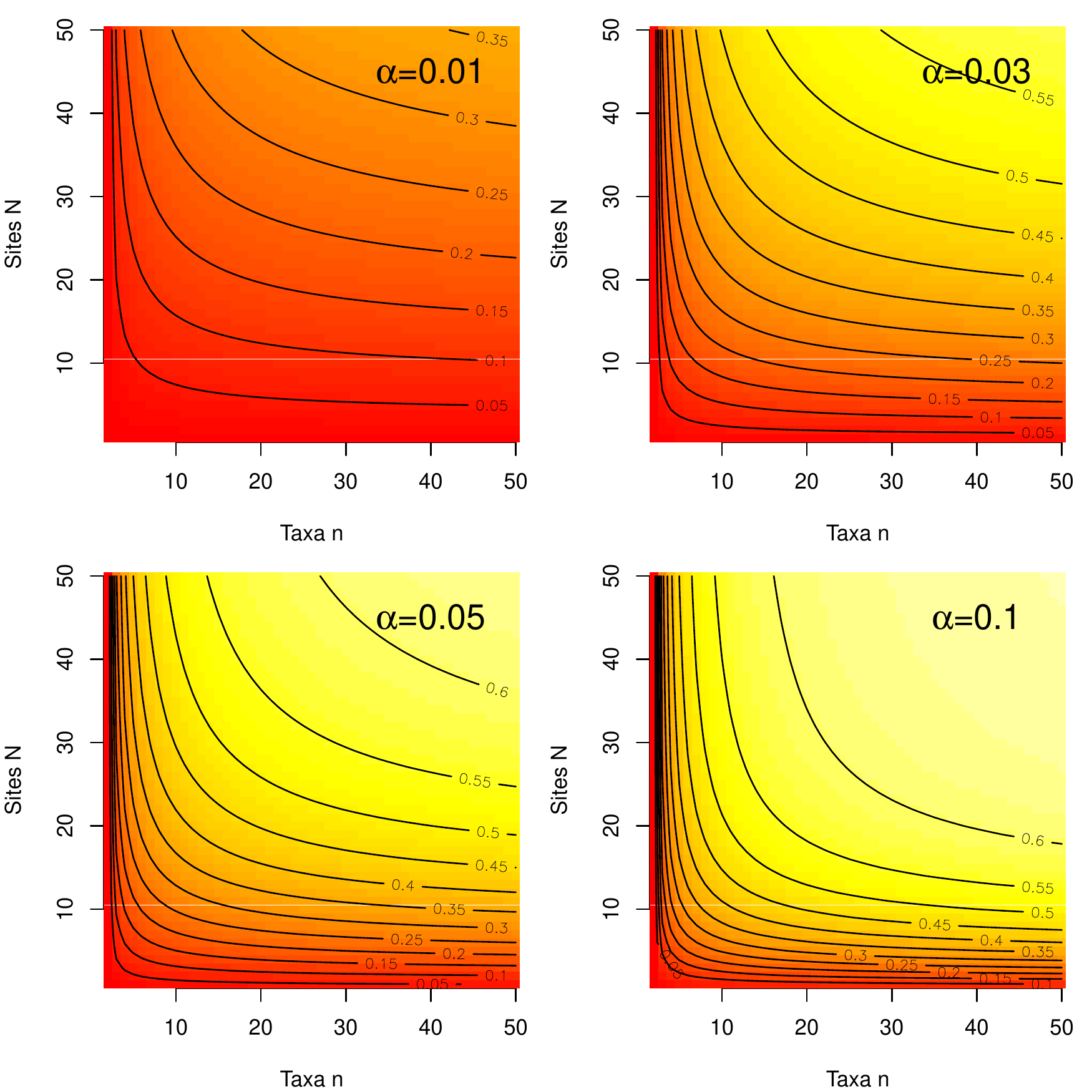}
  \caption{ Comparison of $\widetilde{PI}$ as a function of the number of taxa $n$, number of sites $N$, and mutation rate $\alpha$.  Whether it is better to add taxa or sites to a phylogenetic analysis depends on the mutation rate $\alpha$, and on the number of taxa and sites already present. Greater incremental improvements in $\widetilde{PI}$ per additional taxon or site can be obtained when the mutation rate is larger. }
  \label{fig:pi}
\end{figure}

\citet{Townsend2007Profiling} defines \emph{phylogenetic informativeness} (PI) as the probability that a character evolving according to a Poisson process mutates at least once on a short internal branch of an unrooted four-taxon tree, but does not change again on the terminal branches \citep[See][for details]{Townsend2007Profiling,Townsend2011Taxon}.  The evolutionary rate that maximizes this probability is argued to be optimal for reconstructing the four-taxon tree.  We suggest an extension of this concept to rooted trees with $n$ tips and fixed age $\tau$ as follows.  Suppose we choose two taxa $i$ and $j$, $i\neq j$ at random from the tips of the tree.  

Let $M$ be the number of sites (from $N$ total sites) having at least one mutation shared identically by descent by the chosen taxa under the binary character model in Section \ref{sec:binary}.  This condition corresponds to the event that one or more mutations accumulate on the ancestral branch and none on the two branches leading to randomly chosen taxa $i$ and $j$.  We are interested in the probability of finding at least one site with at least one change.  From \eqref{eq:bink} we have the probability that at least one of $N$ sites has a mutation shared identically by descent,
\begin{equation}
  \Pr(M > 0|x,\alpha,\beta,\text{ibd}) = 1 - \left(1 - P^B_{00}(x)e^{-2\alpha(\tau-x)} - P^B_{01}(x)e^{-2\beta(\tau-x)}\right)^N .
\end{equation}
Now define a measure of phylogenetic informativeness as
\begin{equation}
  \widetilde{\text{PI}} = \sum_{k=1}^{n-1} P(n,k) \int_0^\tau p(x|k,n,\tau,\lambda) \Pr(M>0|x,\alpha,\beta,\text{ibd}) \dx{x} .
    \label{eq:pi}
\end{equation}

An important issue in phylogenetic experimental design is whether to add more taxa or more independent sites to an analysis in order to maximize the chance of accurate tree reconstruction \citep{Sullivan1999Effect,Shpak2000Information,Zwickl2002Increased,Susko2002Testing}.  Our binary mutation model gives one way of answering this question: we seek changes that occur on the ancestral branch and persist identically by descent on both daughter branches.  Figure \ref{fig:pi} shows $\widetilde{\text{PI}}$, given by \eqref{eq:pi}, as a function of the number of sites $N$ and the number of taxa $n$, for different values of the mutation rate $\alpha=\beta$. We set $\tau=1$ and $\lambda=1$ in every case. It is clear from these plots that there is no single answer to the question of whether to add one more site or one more taxon to achieve a unit increase in $\widetilde{PI}$.  When the number of taxa $n$ is small, adding more taxa is best.  When the number of sites $N$ is small, adding more sites is best.  For larger values of the mutation rate $\alpha$, a greater increase in $\widetilde{PI}$ is gained for each taxon or site added.  In cases where there is a significant cost (in money or researcher effort) to obtain an additional sequence sites or taxa, calculation of $\widetilde{PI}$ can help researchers decide whether the additional cost is worth the gain in informativeness.

\subsection{Conditional distribution of MRCA node and age}

It is possible to invert mutation probability expressions to uncover properties of the MRCA of two randomly chosen taxa, conditional on the value of their mutation statistic $M=m$.  In the reversible Poisson model, the probability that the MRCA of two randomly chosen taxa is the $k$th node in the tree is
\begin{equation}
    p(k|\tau,n,m,\lambda,\alpha,\beta) = 
    \frac{\displaystyle  P(n,k) \int_0^\tau p(x|k,n,\tau,\lambda) \Pr(M=m|x,\tau,\alpha,\beta) \dx{x} }{\displaystyle \sum_{j=1}^{n-1} P(n,j) \int_0^\tau p(x|j,n,\tau,\lambda) \Pr(M=m|x,\tau,\alpha,\beta) \dx{x}} .
    \label{eq:condk}
\end{equation}
In a similar way, we can find the conditional distribution of the MRCA age $x$,
\begin{equation}
  p(x|\tau,n,m,\lambda,\alpha,\beta) = \frac{\displaystyle \sum_{k=1}^{n-1} P(n,k)\ p(x|k,n,\tau,\lambda) \Pr(M=m|x,\tau,\alpha,\beta)}{\displaystyle \sum_{k=1}^{n-1} P(n,k) \int_0^\tau p(u|k,n,\tau,\lambda) \Pr(M=m|u,\tau,\alpha,\beta)\dx{u} } .
  \label{eq:condx}
\end{equation}
Given the value of a statistic $M$ which is a function of the trait or character values of two taxa, a rough estimate of the age of their MRCA can be obtained by maximizing \eqref{eq:condx} with respect to $x$.  While this is not a method for tree reconstruction, it may prove useful in settings where only a subset of tip values are observed.  If this subset can be regarded as randomly selected, then measures of evolutionary correlation can still be computed, even in the absence of a full phylogenetic tree.

\section{Discussion}

In this work, we have derived probability distributions for several quantities that give insight into the dynamics of evolutionary processes on unobserved phylogenetic trees.  This is achieved via comparison of evolutionary outcomes for pairs of species.  For continuous trait evolution, we study the evolutionary correlation under Brownian motion and OU processes.  Equation \eqref{eq:bmcorr} provides a natural generalization of the interspecies correlation coefficient $\rho_n$ introduced by \citet{Sagitov2012Interspecies}.  Poisson mutation models with and without reversals provide the distribution of a convenient summary statistic in discrete models of character evolution: the number of changes on an ancestral branch.  The Poisson models also give a natural distribution for the number of segregating sites in an infinite sites model on a tree of age $\tau$ and size $n$.  The distribution of the number of Poisson mutations on the whole tree presented in Corollary \ref{cor:pMpoisson} can also be used to place a posterior distribution on the age $\tau$ of the whole tree.  That is, we can estimate $\tau$, given $M$ observed differences at the tips.  

In the applications presented in Section \ref{sec:applications}, we propose contributions to both prospective and retrospective analysis of evolutionary processes on trees.  First, we extend the notion of phylogenetic informativeness (PI) to trees with $n$ tips and suggest that the choice of whether to add taxa or sites depends on both the number of taxa and mutation rate.  Second, we show that the conditional probability of the MRCA age and location in the tree can be expressed conditional on an observable statistic describing the number of pairwise differences observed.  

A great deal of information about speciation and evolutionary process can be gleaned from pairwise comparisons of taxa.  The simple Yule process provides a parsimonious description of speciation, and makes explicit the correlation induced by the tree in evolutionary outcomes at the tips.  We hope that the results presented in this paper will aid in understanding this correlation and development of inferential techniques for comparative evolutionary analysis.


\section*{Acknowledgements}

We thank Krzysztof Bartoszek for insightful comments and suggestions, and Jeffrey Townsend for helpful conversations about phylogenetic informativeness.


\appendix 


\section{Proof of Theorem \ref{thm:Pnk}}

\label{app:Pnk}

This result is due to \citet{Stadler2009Incomplete}.  Here we present a simple alternative proof based on a recurrence relation.  We provide this derivation because the techniques it employs may be useful in novel derivations of related combinatorial properties of Yule trees, such as those studied by \citet{Mulder2011Probability}.

Let $\nu_n(k)$ denote the number of pairs of tips in an $n$-forest that have the $k$th node of an $n$-tree as their MRCA. We imagine the $n$-forest to be generated from the $(n-1)$-forest by successively grafting one leaf onto each one of the $(n-1)!$ tips, while each time making sure that the newly introduced node is the most recent one on the $n$-tree formed in this manner so that all of them are different.

Suppose $n\geq k+1$. Then $\nu_n(k)$ follows from $\nu_{n-1}(k)$ via induction, based on two observations.  First, any pair of the type considered becomes one in the $n$-forest each time a new tip is attached to one of the $n-3$ tips on the same tree, different from the two members of the pair. Second, when a leaf is added to one of the members of the pair, two new pairs with the property of interest are created, so there are four per pair.  These considerations can be summarised in the recursion relation 
\begin{equation}
 \nu_n(k) = (n-3)\nu_{n-1}(k) + 4\nu_{n-1}(k) = (n+1)\nu_{n-1}(k) .
 \label{eq:nurecur}
\end{equation}
The smallest value of $n$ for which a pair with its MRCA being the $k$th node can occur is $k+1$, and a $(k+1)$-forest contains $\nu_{k+1}(k) = k!$ such pairs (cherries), i.e. exactly one on each tree in the forest.  With \eqref{eq:nurecur}, we obtain $\nu_{k+2}(k) = (k+3)k!$, $\nu_{k+3}(k) = (k+4)(k+3)k!$, etc. In general,
  $\nu_{n}(k) = (n+1)!/(k+1)(k+2)$.
Every choice of picking any pair of tips is equally likely among all $n$-trees whose $k$th node appears between $x$ and $x+\text{d}x$, and since the total number of pairs in the $n$-forest equals $n!(n-1)/2$, the probability of choosing one with the desired property is 
\begin{equation}
  P(n,k) = \frac{2(n+1)}{(n-1)(k+1)(k+2)}
\end{equation}
for $n\geq k+1$, as claimed.


\section{Proof of Lemma \ref{lem:px}}

\label{app:px}

By definition, the first node in the tree has age zero: $p(x|k=1,\tau,n,\lambda) = \delta(x)$, where $\delta(\cdot)$ is the Dirac delta function.  For $k\geq 2$, the $k$th node appears at the moment $k$ lineages become $k+1$.  Suppose the $k$th node appears at time $x+\text{d}x$.  We must first construct a tree beginning with $2$ lineages at time $0$ having $k$ lineages at time $x$, which happens with probability
  $P_{2k}^Y(x) = (k-1) e^{-2\lambda x}(1-e^{-\lambda x})^{k-2}$.
Then at time $x+\text{d}x$, the $k$th node is appears with density $k\lambda e^{-k\lambda \text{d}x}$. Now there are $k+1$ lineages, and the tree must have $n$ lineages at time $\tau$, which happens with probability
\begin{equation}
  P_{k+1,n}^Y(\tau-x) = \binom{n-1}{k}e^{-(k+1)\lambda(\tau-x-\text{d}x)} (1-e^{-\lambda(\tau-x-\text{d}x)})^{n-k-1}
\end{equation}
Finally, since we are conditioning on the full tree having $n$ lineages at time $\tau$, the distribution must be normalised by 
  $P_{2n}^Y(\tau) = (n-1) e^{-2\lambda\tau}(1-e^{-\lambda\tau})^{n-2}$.
Putting these expressions together and sending $\text{d}x$ to zero, we find that 
\begin{equation}
  \begin{split}
    p(x|k,\tau,n,\lambda) &= P_{2k}^Y(x)\ k\lambda\ P_{k+1,n}^Y(\tau-x)/P_{2n}^Y(\tau) \\
                          &= \frac{\lambda (n-2)\binom{n-3}{k-2} e^{-(k-1)\lambda(\tau-x)} (1-e^{-\lambda x})^{k-2} (1-e^{-\lambda(\tau-x)})^{n-k-1}}{(1-e^{-\lambda \tau})^{n-2}}
  \end{split}
\end{equation}
for $k\geq 2$, as claimed.

\section{Proof of Theorem \ref{thm:succinct}}

\label{app:succinct}

With the abbreviation 
  $a = (1-e^{-\lambda x})/(e^{\lambda(\tau-x)}-1)$, 
\eqref{eq:pxn} can be written succinctly in the form 
\begin{equation}
    p(x|n,\tau,\lambda) = \frac{n+1}{3(n-1)}\delta(x) + \lambda \frac{2(n+1)(n-2)(e^{\lambda(\tau-x)}-1)}{(n-1)(1-e^{-\lambda x})^2} \left(\frac{1-e^{-\lambda(\tau-x)}}{1-e^{-\lambda\tau}}\right)^{n-2} \\
  \sum_{k=2}^{n-1} \binom{n-3}{k-2} \frac{a^k}{(k+2)(k+1)} .
  \label{eq:succinct}
\end{equation}
The sum can be evaluated via elementary manipulations to yield
\begin{equation}
  \sum_{k=2}^{n-1} \binom{n-3}{k-2} \frac{a^k}{(k+2)(k+1)}  = \frac{\big((n(n-3)+2) a^2 - 4(n-2)a + 6\big)(1+a)^{n-1} - 2\big((n+1)a + 3\big)}{a^2(n+1)n(n-1)(n-2)} .
\end{equation}
Substitution for $a$ yields \eqref{eq:fullsuccinct}, as claimed. 


\section{Proof of Corollary \ref{cor:pMpoisson}}

\label{app:pMpoisson}

Let 
$R_\tau = \int_0^\tau Y(x)\dx{s} $
be the integral of the Yule process with rate $\lambda$, beginning with $Y(0)=2$ and ending with $Y(\tau)=n$.  Then $R_\tau$ has density function
\begin{equation}
p(x) = \frac{\lambda^{n-2} e^{-\lambda x}}{(n-3)! P_{2n}^Y(\tau)} \sum_{j=2}^n \binom{n-1}{j-1} (j-1) (-1)^{j-2} (x-j \tau)^{n-3}  H(x-j \tau)  
\end{equation}
where $P_{2n}^Y(\tau)$ is the Yule transition probability \eqref{eq:yuleprob} and $H(x)$ is the Heaviside step function \citep{Crawford2013Diversity}.  The minimum branch length that can accrue on the interval $(0,\tau)$ is $2\tau$ and the maximum is $n\tau$.  If the Yule tree has total branch length $x$, then the number of mutations that occur with rate $\alpha$ on the tree has Poisson distribution with rate $\alpha x$; its mass function is $\Pr(M(x) = m|\alpha) = (\alpha x)^m e^{-\alpha x}/m!$.  Then integrating this function with respect to $p(x)$, we have 
\begin{equation}
  \begin{split}
  \Pr(M(\tau)=m|n,\tau,\lambda,\alpha) &= \int_0^\infty \Pr(M(x)=m|\alpha) \ p(x) \dx{x}\\
  &= \int_{2\tau}^{n\tau} \frac{(\alpha x)^m e^{-\alpha x}}{m!}  \frac{\lambda^{n-1} e^{-\lambda x}}{(n-2)!} \sum_{j=1}^n \binom{n-1}{j-1} (-1)^{j-1} (x-j \tau)^{n-2}  H(x-j \tau)   \dx{x}\\
                  &= \frac{\alpha^m\lambda^{n-1}}{m!(n-2)!} \sum_{j=1}^n \binom{n-1}{j-1} \sum_{k=0}^{n-2} \binom{n-2}{k} (-1)^{n-k+j-3}(j\tau)^{n-k-2} \int_{j\tau}^{n\tau} x^{k+m} e^{-(\alpha+\lambda)x} \dx{x}\\
  \end{split}
\end{equation}
Re-writing the integral as the difference of lower incomplete Gamma functions delivers the desired expression.


\section{Proofs of \eqref{eq:poisPm0} and \eqref{eq:poisPm}}

\label{app:poisPm}

  We seek an expression for the probability that $m$ mutations accumulate on the ancestral branch of two randomly chosen taxa,
  \begin{equation}
    \Pr(M=m|n,\tau,\lambda,\alpha) = \sum_{k=2}^{n-1} P(n,k) \int_0^\tau p(x|k,n,\tau,\lambda) \frac{(\alpha x)^m e^{-\alpha x}}{m!} \dx{x}.
  \end{equation}
  Suppose for now that $M\geq 1$.  Then the Poisson mutation probability is zero when the MRCA node $k=1$, and the first term in the sum drops out.  Marginalizing over $x$ and $k$, we find 
\begin{equation}
  \begin{split}
    \Pr(M=m|n,\tau,\lambda,\alpha) &= \frac{2\lambda(n+1)(n-2)\alpha^m}{m!(n-1)(1-e^{-\lambda \tau})^{n-2}}\sum_{k=2}^{n-1} \binom{n-3}{k-2} \frac{e^{-(k-1)\lambda \tau} }{(k+1)(k+2)} \\
                           &\quad \times \int_0^\tau e^{(\lambda(k-1)-\alpha)x} x^m (1-e^{-\lambda x})^{k-2} (1-e^{-\lambda(\tau-x)})^{n-k-1} \dx{x} \\
                           &= \frac{2\lambda(n+1)(n-2)\alpha^m}{m!(n-1)(1-e^{-\lambda\tau})^{n-2}}\sum_{k=2}^{n-1} \binom{n-3}{k-2} \frac{e^{-(k-1)\lambda\tau} }{(k+1)(k+2)} \\
                           &\quad \times \int_0^\tau e^{(\lambda(k-1)-\alpha)x}  x^m \left[ \sum_{j=0}^{k-2} \binom{k-2}{j} (-1)^j e^{-\lambda j x}\right] \left[ \sum_{i=0}^{n-k-1} \binom{n-k-1}{i} (-1)^i e^{-\lambda i(\tau-x)}\right] \dx{x} \\
                           &= \frac{2\lambda(n+1)(n-2)\alpha^m}{m!(n-1)(1-e^{-\lambda\tau})^{n-2}}\sum_{k=2}^{n-1} \binom{n-3}{k-2} \frac{e^{-\lambda (k-1)\tau} }{(k+1)(k+2)} \\
                           &\quad \times \sum_{j=0}^{k-2} \binom{k-2}{j} \sum_{i=0}^{n-k-1} \binom{n-k-1}{i} (-1)^{i+j} e^{-i\lambda\tau} \int_0^\tau x^m \exp\left[ -x (\alpha+\lambda(j-i-k+1))\right]\dx{x} .
\end{split}
\end{equation}
Writing the integral as an incomplete gamma function gives the result for $M\geq 1$.  Now consider the case that $M=0$.  When the MRCA of the two chosen taxa is $k=1$, the time of shared ancestry is $x=0$, and no mutations can accumulate.  Therefore 
  \begin{equation}
    \begin{split}
      \Pr(M=0|n,\tau,\lambda,\alpha) &= \frac{n+1}{3(n-1)} + \sum_{k=2}^{n-1} P(n,k) \int_0^\tau p(x|k,n,\tau,\lambda) e^{-\alpha x} \dx{x}  \\
    &= \frac{n+1}{3(n-1)} + \frac{2(n+1)(n-2)}{(n-1)(1-e^{-\lambda\tau})^{n-2}}\sum_{k=2}^{n-1} \binom{n-3}{k-2} \frac{e^{-(k-1)\lambda\tau} }{(k+1)(k+2)} \\
                           &\quad \times \sum_{j=0}^{k-2} \binom{k-2}{j} \sum_{i=0}^{n-k-1} \binom{n-k-1}{i} (-1)^{i+j} e^{-i\lambda\tau} \int_0^\tau x^m \exp\left[ -x (\alpha+\lambda(j-i-k+1))\right]\dx{x} .
  \end{split}
  \end{equation}
  Again replacing the integral by the incomplete gamma function gives the result.


\section{Proof of \eqref{eq:poisrevPm}}
\label{app:poisrevPm}

Since all mutations are guaranteed to be unique, shared mutations must have arisen during the time of shared ancestry $x$.  Likewise, mutations that arise on independent branches after the MRCA of the two taxa are guaranteed to be unique, so we do not need to account for them in the shared total.  In order for the two taxa to share exactly $m$ mutations, some number $k\geq m$ mutations must have occurred during the time $x$ of shared ancestry (given by \eqref{eq:poisrev}), and then some set of $m$ mutations must be preserved and not lost during $\tau-x$.  The distribution of the number of mutations on the ancestral branch is Poisson with rate $\frac{\alpha}{\beta}(1-e^{-\beta x})$
and the probability of $m$ surviving in both lineages during $\tau-x$ is 
\begin{equation}
\binom{j}{m} e^{-2\beta m(\tau-x)} (1-e^{-2\beta(\tau-x)})^{j-m} \\
\end{equation}
Then marginalizing over the number $j$ of mutations at the splitting time $x$, we have 
\begin{equation}
  \begin{split}
    \Pr(M=m|x,\tau,\alpha,\beta) &= \sum_{j=m}^\infty \frac{\left(\frac{\alpha}{\beta}(1-e^{-\beta x})\right)^j \exp\left[-\frac{\alpha}{\beta}(1-e^{-\beta x})\right]}{j!} \binom{j}{m} e^{-2\beta m(\tau-x)} (1-e^{-2\beta(\tau-x)})^{j-m} \\
    &= \frac{\exp\left[-\frac{\alpha}{\beta}(1-e^{-\beta x})\right]}{m!} \left[ \frac{\alpha}{\beta}(1-e^{-\beta x}) e^{-2\beta(\tau-x)} \right]^m \sum_{j=m}^\infty \frac{[\frac{\alpha}{\beta}(1-e^{-\beta x})(1-e^{-2\beta(\tau-x)})]^{j-m}}{(j-m)!}  \\
    &= \frac{\exp\left[-\frac{\alpha}{\beta}(1-e^{-\beta x})\right]}{m!} \left[ \frac{\alpha}{\beta}(1-e^{-\beta x}) e^{-2\beta(\tau-x)} \right]^m \exp\left[\frac{\alpha}{\beta}(1-e^{-\beta x})(1-e^{-2\beta(\tau-x)})\right] \\
    &= \frac{\exp\left[-\frac{\alpha}{\beta}(1-e^{-\beta x})e^{-2\beta(\tau-x)} \right] \left[ \frac{\alpha}{\beta}(1-e^{-\beta x}) e^{-2\beta(\tau-x)} \right]^m }{m!}
  \end{split}
\end{equation}
as claimed.



\section*{References}

\bibliographystyle{spbasic}
\bibliography{mutationcount}

\end{document}